\documentclass[12pt]{article}

\title{Test of a theoretical equation of state for elemental solids 
       and liquids}

\author{Eric D.\ Chisolm, Scott D.\ Crockett, and Duane C.\ Wallace \\ 
        Theoretical Division, Los Alamos National Laboratory \\ 
        Los Alamos, NM~~87545}

\usepackage{graphicx}

\begin{document}

\maketitle

\vspace*{-3.0in} \begin{flushright} LA-UR-03-3264 \end{flushright}
\vspace*{2.5in}

\begin{abstract}
We propose a means for constructing highly accurate equations of state
(EOS) for elemental solids and liquids essentially from first
principles, based upon a particular decomposition of the underlying
condensed matter Hamiltonian for the nuclei and electrons.  We also
point out that at low pressures the neglect of anharmonic and
electron-phonon terms, both contained in this formalism, results in
errors of less than 5\% in the thermal parts of the thermodynamic
functions.  Then we explicitly display the forms of the remaining terms
in the EOS, commenting on the use of experiment and electronic
structure theory to evaluate them.  We also construct an EOS for
Aluminum and compare the resulting Hugoniot with data up to 5 Mbar,
both to illustrate our method and to see whether the approximation of
neglecting anharmonicity et al.\ remains viable to such high
pressures.  We find a level of agreement with experiment that is
consistent with the low-pressure results.
\end{abstract}

\section{Introduction}
\label{intro}

Over approximately the last sixty years, numerous models and
techniques have been developed for creating equations of state (EOS)
for a variety of materials that are valid up to very extreme pressures
(tens of Mbar) and temperatures (several eV).  In the EOS community at
the national laboratories, for instance, we have often used models
based on the Mie-Gr\"{u}neisen EOS together with the Thomas-Fermi or
Thomas-Fermi-Dirac model (or one of its modifications) to include the
contributions from the electrons (see \cite{manual} for examples).
The models usually contain enough independent parameters to adjust the
EOS until it correctly reproduces the experimentally measured Hugoniot
(and perhaps a few other data points), but it is generally an open
question how accurate the EOS is away from the Hugoniot.  In this
paper we argue that for one class of materials, elemental solids and
liquids, our understanding of the underlying condensed matter
Hamiltonian for the nuclei and electrons has grown to the point that
we can construct highly accurate EOS from essentially first
principles, and we also propose a means for doing so.  We also argue
that, since the underlying physics is well understood, an EOS derived
this way should have the right functional form, even if we are unsure
of the values of some of its parameters; thus, if the resulting EOS is
shown to be accurate in one thermodynamic region (say, along the
Hugoniot), then we can be confident that it is roughly equally
accurate elsewhere.

In this formalism, the EOS in the solid phase depends on a
decomposition of the Hamiltonian due to Wallace (see Chapter 1 of
\cite{book}), extending the work of Born \cite{born} to metals as well
as insulators; the resulting free energy contains terms describing the
harmonic motion of the nuclei about their lattice sites (phonons),
thermal excitation of the electrons from their ground state,
anharmonic corrections to the nuclear motion (represented as
phonon-phonon interactions), and interactions between the electron
excitations and the nuclear motion, represented as electron-phonon
interactions.  (Please note that this description is exact; all of the
physics contained in the true Hamiltonian of the system is included
here.  Specific EOS models usually neglect the anharmonic and
electron-phonon terms, arguing that anharmonicity is small and making
reference to some form of the Born-Oppenheimer approximation; we will
take a somewhat different route, commenting on approximations below.)
A recently developed theory of the dynamics of monatomic liquids (see
\cite{liq} for a review) uses the same Hamiltonian to derive a liquid
free energy which is quite similar to the expression for a solid, with
additional terms accounting for the fact that the liquid, as opposed
to the solid, traverses many potential valleys and thus sees the
boundaries between them.  For both phases, the resulting free energies
have been compared with experimental data in the low-pressure regime
($P \leq 100$ kbar), with the following results (Sections 17-19 and 23 
of \cite{book}):

(a) Molecular dynamics (MD) calculations of the anharmonic
contribution to the entropy of several solids match experimental
entropy data to the accuracy of the data themselves.

(b) Low-temperature ($T \leq 20$ K) calculations of the
electron-phonon term for several solids lead to predictions that also
match experimental entropy to the accuracy of the data.

(c) Theoretical arguments show that the electron-phonon contribution
is entirely negligible except when the electronic contribution
dominates the free energy, such as in metallic solids at low
temperatures.

(d) For the 27 elemental solids for which accurate data are available
from low $T$ (but not too low; see point (c)) to the melting
temperature $T_m$, the free energy excluding the anharmonic and
electron-phonon terms accounts for the experimental thermal energy and
entropy to an accuracy of $5\%$ (in fact, an accuracy of $2\%$ for all
but about five materials).

(e) For the 6 elements in the liquid phase for which accurate data are
available at temperatures up to around $3\,T_m$, the effect of
neglecting the anharmonic, boundary, and electron-phonon contributions
to the energy and entropy is similarly small.

This tells us that at low pressures, we can neglect the anharmonic,
boundary, and electron-phonon terms in both the solid and liquid free
energy (which happen to be the hardest terms to calculate), and the
resulting thermal energy and entropy are both simple in form and
accurate at the $5\%$ level.  It is for this reason, not an appeal to
the Born-Oppenheimer or other approximations, that we know we can
simplify our EOS and what the results of the simplification will be,
at least at low pressures.

In this paper we do two things: (1) We describe in more detail this
framework for constructing EOS and discuss the theoretical and
experimental inputs needed to implement it, and (2) we construct a
sample EOS, neglecting anharmonic, boundary, and electron-phonon
terms, both to illustrate the method and to discover whether points
(d) and (e) above continue to hold in the high-pressure regime.  We
use Aluminum as our sample because of the availability of extensive
electronic structure calculations, up to a compression of three, and
highly accurate shock Hugoniot data, which provide a test of our EOS
through both phases to pressures of around 5 Mbar.  In Subsection
\ref{solid} we develop the general theory of the solid EOS, and in
Subsection \ref{liquid} we do the same for the liquid.  In Subsection
\ref{constructing}, we construct our sample EOS for Al, comparing it
with other EOS work, and in Subsection \ref{hug} we compute the
Hugoniot predicted by the EOS and compare it with experimental data.
The results are encouraging.  Finally, we review our work, discuss the
advantages and disadvantages of this formalism (and how to address the
disadvantages), and suggest directions for future development.

\section{General theory}
\label{theory}

\subsection{Solid phase}
\label{solid}

The condensed matter Hamiltonian, decomposed as described above,
consists of terms describing the motion of the nuclei in a potential
generated by the electrons in their ground state, plus additional
terms that lead to the thermal excitation of the electrons and
describe their interactions with the nuclear motion.  With this
Hamiltonian, the Helmholtz free energy per atom for a solid at
temperature $T$ with volume $V$ per atom takes the form
\begin{equation} 
F^{\rm s}(V,T) = \Phi_0^{\rm s}(V) + F_{\rm ph}^{\rm s}(V,T) + 
F_{\rm el}^{\rm s}(V,T) + F_{\rm anh}^{\rm s}(V,T) + F_{\rm ep}^{\rm s}(V,T).
\label{Fs}
\end{equation}
Here $\Phi_0^{\rm s}$ is the static lattice potential (the electronic
ground state energy when the nuclei are fixed at their lattice sites);
it depends on the particular crystal structure.  $F_{\rm ph}^{\rm s}$
is the contribution from the harmonic motion of the nuclei about their
lattice sites, $F_{\rm el}^{\rm s}$ represents the thermal excitation
of the electrons when the nuclei are fixed at their lattice sites,
$F_{\rm anh}^{\rm s}$ accounts for the anharmonicity of the nuclear
motion (which may be represented as phonon-phonon interactions), and
$F_{\rm ep}^{\rm s}$ expresses the interactions between the electron
excitations and the nuclear motion, represented as electron-phonon
interactions.  (We emphasize again that this free energy is exact; it
includes all of the physics present in the Hamiltonian.)  The
discussion in the Introduction justifies our approximating the solid free
energy as
\begin{equation} 
F^{\rm s}(V,T) = \Phi_0^{\rm s}(V) + F_{\rm ph}^{\rm s}(V,T) + 
F_{\rm el}^{\rm s}(V,T),
\label{realFs}
\end{equation}
so let us now consider the forms of $F_{\rm ph}^{\rm s}$ and $F_{\rm
el}^{\rm s}$ and the parameters on which they depend.  The phonon term
in the Hamiltonian describes harmonic motion, which leads uniquely to
the free energy of lattice dynamics:
\begin{equation}
F_{\rm ph}^{\rm s}(V,T) = \int_0^\infty g^{\rm s}(\omega) \left[
\frac{1}{2}\hbar\omega + \ln(1-e^{-\beta\hbar\omega}) \right] d\omega,
\label{Flatt}
\end{equation}
where $\beta = 1/kT$ and $g^{\rm s}(\omega)$ is the distribution of
phonon frequencies in the Brillouin zone.  (Note that $g^{\rm s}
(\omega)$ is volume dependent.)  Sometimes we require not the full
Eq.\ (\ref{Flatt}) but only its high- and low-temperature limits, for
which we need not the full $g^{\rm s}(\omega)$ but only three of its
moments, expressed in terms of the characteristic temperatures
$\Theta_0^{\rm s}$, $\Theta_1^{\rm s}$, and $\Theta_2^{\rm s}$ defined
by
\begin{eqnarray}
\ln k\Theta_0^{\rm s} & = & \langle \ln \hbar \omega \rangle_{\rm BZ} 
                            \nonumber \\
k\Theta_1^{\rm s} & = & \frac{4}{3} \langle \hbar \omega \rangle_{\rm BZ} 
                  \nonumber \\
k\Theta_2^{\rm s} & = & \left[ \frac{5}{3} \langle (\hbar \omega)^2 
                        \rangle_{\rm BZ} \right]^{1/2},
\end{eqnarray}
where $\langle \cdots \rangle_{\rm BZ}$ indicates an average over all
the frequencies in the Brillouin zone.  Then the following limits hold:
\begin{equation}
F_{\rm ph}^{\rm s}(V,T) \rightarrow \frac{9}{8}k\Theta_1^{\rm s} 
  {\rm \ \ \ as \ \ \ } T \rightarrow 0
\label{FlattlowT}
\end{equation}
and
\begin{equation}
F_{\rm ph}^{\rm s}(V,T) = -3kT\left[\ln\left(\frac{T}{\Theta_0^{\rm s}}\right) 
      - \frac{1}{40}\left(\frac{\Theta_2^{\rm s}}{T}\right)^2 + \cdots \right] 
               {\rm \ \ at \ high \ } T.
\label{FlatthighT}
\end{equation}
The leading term in Eq.\ (\ref{FlatthighT}) describes purely classical
nuclear motion, while the series of terms in powers of $T^{-2}$ are
quantum corrections.  Keeping only the first quantum correction, the
thermodynamic functions derived from Eq.\ (\ref{FlatthighT}) are
accurate to $1\%$ at temperatures above $\frac{1}{2}\Theta_2^{\rm s}$.

The electronic excitation free energy $F_{\rm el}^{\rm s}$ can be
expressed generally as an integral function of the electronic density
of states per atom, $n^{\rm s}(\epsilon)$, and the Fermi distribution
\begin{equation}
f(\epsilon) = \frac{1}{e^{\beta(\epsilon - \mu)} + 1},
\end{equation}
where $\beta$ is still $1/kT$ and $\mu$ is the chemical potential.  If
each atom contributes $Z$ electrons to the valence bands (notice that
$Z$ is not necessarily the atomic number), with the lowest valence
energy set to zero, then $\mu$ is a function of $T$ determined by the
normalization
condition
\begin{equation} 
\int^\infty_0 n^{\rm s}(\epsilon) f(\epsilon)\, d\epsilon = Z.
\label{findmu}
\end{equation} 
The electronic free energy is then
\begin{eqnarray}
\lefteqn{F_{\rm el}^{\rm s}(V,T) = } \nonumber \\
& & \mu Z - \int^{\epsilon_F}_0 \epsilon \, n^{\rm s}(\epsilon) \, d\epsilon - 
    kT\int^\infty_0 n^{\rm s}(\epsilon) \ln[1+e^{-\beta(\epsilon - \mu)}]\, 
    d\epsilon,
\label{Felgen}
\end{eqnarray}
where $\epsilon_F$, the Fermi energy, is the value of $\mu$ when
$T=0$.  The second term on the right hand side of Eq.\ (\ref{Felgen})
is the subtraction of the electronic ground state energy, which
ensures that $F_{\rm el}^{\rm s} \rightarrow 0$ as $T \rightarrow 0$.
This property makes sense if $F_{\rm el}^{\rm s}$ represents purely
thermal excitation of the electrons.  (It also avoids double counting
of the energy, as the electronic ground state energy is already
represented as $\Phi_0^{\rm s}$.)

We see from this discussion that to evaluate the terms in Eq.\
(\ref{realFs}) for the solid free energy we require three unknown
functions: $\Phi_0^{\rm s}$, $g^{\rm s}(\omega)$ (or $\Theta_0^{\rm
s}$, $\Theta_1^{\rm s}$, and $\Theta_2^{\rm s}$ if we are concerned
only with the high- and low-$T$ limits), and $n^{\rm s}(\epsilon)$
(and the associated quantities $Z$ and $\epsilon_F$).  These can be
determined in various ways: compressibility data and diamond anvil
cell data can be used to construct $\Phi_0^{\rm s}(V)$; neutron
scattering experiments can determine $g^{\rm s}(\omega)$ or its
various moments at $P = 1$ bar; and for many elements all three of
these functions can be computed reliably using electronic structure
theory.  (Or one could use results from multiple sources in
combination, which is often an option with $\Phi_0^{\rm s}$ and is
basically a necessity with $g^{\rm s}(\omega)$.)  One must keep in
mind, however, that the accuracy of one's answers will be limited by
the accuracy and range of applicability of these functions, regardless
of how they are determined.

\subsection{Liquid phase and two-phase region}
\label{liquid}

According to the theory of liquid dynamics reviewed in \cite{liq}, the
same Hamiltonian that gave us the solid free energy leads to a similar
form for the free energy of a monatomic liquid.  In this theory, the
region of the many-body potential surface in which the system moves in
the liquid phase is dominated by a large number of intersecting
nearly-harmonic valleys, called ``random'' valleys because they
correspond to particle configurations which retain no remnant crystal
symmetry, and which are all macroscopically identical.  In particular,
the valleys all have the same distribution of normal mode frequencies,
and they all have the same depth (which, as in the solid case, is the
electronic ground state energy when the nuclei are fixed at the valley
minimum).  The resulting liquid free energy per atom is
\begin{eqnarray} 
F^l(V,T) & = & \Phi_0^l(V) + F_{\rm ph}^l(V,T) + 
F_{\rm el}^l(V,T) + F_{\rm ab}^l(V,T) +  \nonumber \\ 
& & F_{\rm ep}^l(V,T) - kT\ln w.
\label{Fl}
\end{eqnarray}
All of the terms correspond to their solid counterparts with the
following exceptions:

(1) $\Phi_0^l$, now called the static {\em structure} potential, is
the depth of a typical valley in which the liquid system moves.

(2) The normal mode spectrum appearing in $F_{\rm ph}^l$ is that
of a typical liquid potential valley, not the unique solid potential
valley.

(3) The term $F_{\rm ab}^l$ includes corrections due both to
anharmonicity and to the fact that the potential valleys have
boundaries, which the liquid (as opposed to the solid) encounters as
it transits from valley to valley.

(4) The extra term $-kT\ln w$ corresponds to an increase in entropy of
$k\ln w$ per atom; the value $\ln w \approx 0.8$ is estimated from
entropy of melting data of the elements (again, see \cite{liq} for
details).  In liquid dynamics theory, this term is due to the
hypothesis that the number of potential valleys among which the liquid
moves is of order $w^N$, where $N$ is the number of atoms in the
system.

We emphasize that the same Hamiltonian gives rise to both Eq.\
(\ref{Fs}) and (\ref{Fl}); the differences are that the potential is
expanded about different equilibrium configurations in the two cases,
and that the region of configuration space over which the liquid moves
is obviously far larger than the space available to the solid (hence
the $-kT\ln w$ term).

Again making the approximations discussed in the Introduction, our
form for the liquid free energy becomes
\begin{equation}
F^l(V,T) = \Phi_0^l(V) + F_{\rm ph}^l(V,T) + F_{\rm el}^l(V,T) - kT\ln w,
\label{realFl}
\end{equation}
and the additional term $-kT\ln w$ is fully determined by setting $\ln
w = 0.8$, as mentioned above.  The form of the phonon term is dictated
by a central hypothesis of liquid dynamics theory: The motion of the
liquid consists of oscillations in the macroscopically similar valleys
described above together with occasional {\em transits} between
valleys; the transits are of such short duration that they do not
contribute to the thermodynamics to lowest order.  Thus we will take
$F_{\rm ph}^l$ to have the same form as the solid phonon term, Eq.\
(\ref{Flatt}), with a possibly different phonon frequency distribution
$g^l(\omega)$.  The electronic excitation term for the solid was
derived using only the assumption that the electrons are thermally
distributed over the available states using Fermi statistics; all of
the information about the configuration of the nuclei is contained in
the density of states.  Hence $F_{\rm el}^l$ also takes the same form
as the corresponding solid term, Eq.\ (\ref{Felgen}), with a density
of states $n^l(\epsilon)$ appropriate for the liquid phase.  (What
this means is discussed briefly below.)

The liquid and solid EOS together determine the melting temperature
as a function of pressure $T_m(P)$ by the requirement that the solid
and liquid Gibbs free energies match along the melt curve, or
\begin{equation}
G^{\rm s}(P, T_m(P)) = G^l(P, T_m(P)).
\label{Gibbsmatch}
\end{equation}
Once the melt curve is determined, one can use the solid and liquid
EOS separately to compute $V^{\rm s}_m(T)$ and $V^l_m(T)$, the atomic
volumes of the solid and liquid at melt as functions of temperature.
(Of course, using $T_m(P)$ one can express $V^{\rm s}_m$ and $V^l_m$
as functions of pressure equally well.)

In the case $V^{\rm s}_m(T) < V^l_m(T)$ for all $T$, which we assume
here, the computation of the full two-phase EOS ($F$, $E$, $S$, and
$P$) for any $V$ and $T$ proceeds as follows.  If $V \leq V^{\rm
s}_m(T)$ for the given $T$, the system is in the solid phase; $F^{\rm
s}$ is computed as described in Subsection \ref{solid}, and the other
functions follow from
\begin{eqnarray}
S^{\rm s} & = & -\left(\frac{\partial F^{\rm s}}{\partial T}\right)_V, 
                \nonumber \\  
E^{\rm s} & = & F^{\rm s} + TS^{\rm s}, \nonumber \\
P^{\rm s} & = & -\left(\frac{\partial F^{\rm s}}{\partial V}\right)_T. 
\label{thermfuncs}
\end{eqnarray}
If $V \geq V^l_m(T)$, the system is in the liquid phase; $F^{\rm l}$
is computed as described in this Subsection, and the other functions
follow from expressions analogous to Eqs.\ (\ref{thermfuncs}).  If
$V^{\rm s}_m(T) < V < V^l_m(T)$, the system is in the two-phase
region; defining $\eta$, the volume fraction of the system in the
liquid phase, by
\begin{equation} 
\eta = \frac{V - V^{\rm s}_m(T)}{V^l_m(T) - V^{\rm s}_m(T)}, 
\end{equation}
the thermodynamic functions are
\begin{eqnarray}
F(V,T) & = & \eta \, F^l(V^l_m(T),T) + 
               (1-\eta) F^{\rm s}(V^{\rm s}_m(T),T), \nonumber \\
E(V,T) & = & \eta \, E^l(V^l_m(T),T) + 
               (1-\eta) E^{\rm s}(V^{\rm s}_m(T),T), \nonumber \\
S(V,T) & = & \eta \, S^l(V^l_m(T),T) + 
               (1-\eta) S^{\rm s}(V^{\rm s}_m(T),T), \nonumber \\
P(V,T) & = & -\,\frac{F^l(V^l_m(T),T) - 
                     F^{\rm s}(V^{\rm s}_m(T),T)}
                    {V^l_m(T) - V^{\rm s}_m(T)}.   
\end{eqnarray}

Finally, we note that just as with the solid, to evaluate the terms in
Eq.\ (\ref{realFl}) we require three unknown functions: $\Phi_0^l$,
$g^l(\omega)$ (or $\Theta_0^l$ and $\Theta_2^l$, since we're not
likely to need the low-$T$ limit), and $n^l(\epsilon)$.  In contrast
to the solid case, these functions are generally not available
experimentally.  (It is possible that one might be able to compute
$\Phi_0^l$ using liquid compressibility data, but we suspect that this
will be very difficult.)  However, for many materials these functions
should be computable using electronic structure theory, proceeding
much as one would in the solid case except that the nuclei would be
arranged not in a crystal configuration but in a disordered structure
characteristic of a ``random'' valley in the liquid potential surface
\cite{liq}.  To our knowledge very few such calculations have been
attempted; the only ones we are aware of are $\Phi_0^l$ and
$g^l(\omega)$ at a single volume for liquid sodium in \cite{sod1} (the
results are referred to in \cite{liq} and a graph of $g^l(\omega)$
using their results appears as Fig.\ 1 in \cite{sod2}).  Another
function that is sometimes available is the melt curve $T_m(P)$, but
this curve cannot be chosen independently of the others, since the
solid and liquid EOS determine it jointly; but this can be an
advantage, since if $T_m(P)$ is known from experiment, for example, it
can be used to compute one of the other needed functions if it is not
otherwise available.  In fact, this is how we will determine
$\Phi_0^l$ in our example EOS, to which we now turn.

\section{An example: Aluminum}
\label{Al}

To illustrate the application of the theory we've described, we will
now construct an EOS for Aluminum, which has been the subject of
extensive electronic structure calculations and for which a great deal
of high-quality experimental data are available.  We will then compare
the Hugoniot predicted by our EOS with data up to pressures of
approximately 5 Mbar; this will tell us whether the approximations we
discussed in the Introduction (neglecting anharmonic, boundary, and
electron-phonon effects), known to be accurate at low pressures,
continue to be reasonable in the high-pressure domain.

\subsection{Constructing the EOS}
\label{constructing}

We recall from Subsection \ref{solid} that the solid EOS requires
three functions: $\Phi_0^{\rm s}$, $g^{\rm s}(\omega)$, and $n^{\rm
s}(\epsilon)$.  Since we will be testing the EOS by comparison with
Hugoniot data, we will always be in the high-$T$ region (except for
one brief low-$T$ excursion; see below), so we use Eq.\
(\ref{FlatthighT}) for $F^{\rm s}_{\rm ph}$ instead of Eq.\
(\ref{Flatt}); this means that we require only $\Theta_0^{\rm s}$,
$\Theta_1^{\rm s}$, and $\Theta_2^{\rm s}$ in place of $g^{\rm
s}(\omega)$.  

To determine these functions, we began by consulting the results of
density functional theory (DFT) calculations carried out in the local
density approximation by Straub et al.\ \cite{straub}.  They worked
with fcc and bcc Al at atomic volumes from $37\ a_0^3$ to $160\
a_0^3$, where $a_0$ is the Bohr radius, corresponding to densities
from $8.17$ g/cm$^3$ to $1.89$ g/cm$^3$ (the density of Al at 293 K
and 1 bar is $2.700$ g/cm$^3$).  Their calculations indicate a $T=0$
transition from fcc to bcc at $51\ a_0^3$, corresponding to
$\rho=5.93$ g/cm$^3$, but we will neglect this phase change and treat
solid Al as an fcc crystal for two reasons: The DFT calculations
themselves suggest that the effect of the phase change on the
thermodynamic functions will be quite small; and we know from
experiment that the solid-liquid transition on the Hugoniot takes
place well before reaching the density of concern (see Subsection
\ref{hug}), so we are confident of our assumption of fcc along the
Hugoniot until melting.  However, this assumption may have an effect on
the liquid EOS at high densities, which we will comment on below.
(Other electronic structure work, discussed on pp.\ 89-90 of
\cite{young}, suggests the possibility of an hcp phase between the fcc
and bcc phases, but as \cite{young} also mentions, no experimental
signature of this phase has been found, so we will proceed under the
assumption of a single solid phase.)

\begin{figure}
\centering
\includegraphics[scale=0.5]{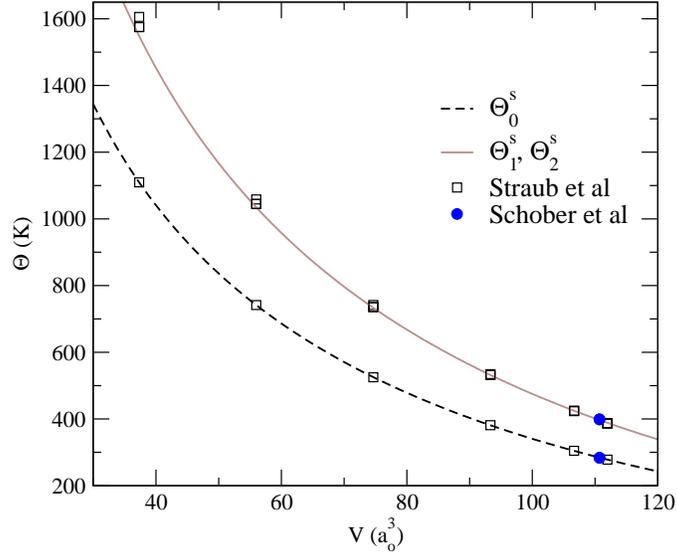}
\caption{$\Theta_0^{\rm s}$, $\Theta_1^{\rm s}$, and $\Theta_2^{\rm
s}$ as functions of atomic volume from the DFT calculations in
\cite{straub} and experimental data from \cite{schob}.  Our functional
forms are also shown.}
\label{Thetas}
\end{figure}

Straub et al.\ computed $\Phi_0^{\rm s}$ for fcc by fitting their
results to a Birch-Murnaghan form,
\begin{equation}
\Phi_0^{\rm s}(V) = c_0 + V_b \sum_{n=2}^{5} \frac{c_n}{n!} \left\{ 
\frac{1}{2} \left[\left(\frac{V}{V_b} \right)^{-2/3} - 1 \right]\right\}^n,
\label{firstphi}
\end{equation}
with coefficients
\begin{eqnarray}
& & V_b = 106.302\,a_0^3, \ \ \ \ \ \ \ \ \ c_0 = -287.7832\ {\rm mRy}, 
\nonumber \\
& & c_2 = 761.2029\ {\rm GPa}, \ \ \ \ \, c_3 = 1319.036\ {\rm GPa}, 
\nonumber \\
& & c_4 = -13\,661.06\ {\rm GPa}, \ \, c_5 = 50\,315.53\ {\rm GPa}.
\end{eqnarray}
The $\Theta_n^{\rm s}$ were determined by computing the bulk modulus
and four zone-boundary phonons at several volumes, and these results
were used to calibrate a pseudopotential model at each volume.  The
pseudopotential was then used to calculate phonon frequencies
throughout the Brillouin zone, allowing the determination of
$\Theta_0^{\rm s}$, $\Theta_1^{\rm s}$, and $\Theta_2^{\rm s}$.  Their
results are shown in Table \ref{Thetatab} and Figure \ref{Thetas}.
(The full set of results was not reported in \cite{straub}.)
\begin{table}
\centering
\begin{tabular}{c|ccc} \hline
$V$ ($a_0^3$) & $\Theta_0^{\rm s}$ (K) & $\Theta_1^{\rm s}$ (K) & 
                                $\Theta_2^{\rm s}$ (K) \\ \hline
111.97 & 278.09 & 386.55 & 387.20 \\
106.65 & 304.63 & 423.81 & 424.86 \\
93.318 & 381.43 & 532.00 & 534.48 \\
74.655 & 525.01 & 735.49 & 741.68 \\
55.991 & 741.62 & 1044.7 & 1058.3 \\
37.327 & 1109.5 & 1575.0 & 1605.4 \\ \hline
\end{tabular}
\caption{DFT calculations of $\Theta_0^{\rm s}$, $\Theta_1^{\rm s}$,
and $\Theta_2^{\rm s}$ from \cite{straub}.}
\label{Thetatab}
\end{table}
To check these results, Straub et al.\ compared experimental phonon
moments for Al at $T=80$ K and $P=1$ bar based on Born-von Karmen fits
to neutron scattering data \cite{schob} with their predictions
interpolated to the appropriate atomic volume of $110.7\ a_0^3$.  The
experimental points, also shown in Figure \ref{Thetas}, are in very
good agreement with their calculations; hence these results for the
$\Theta_n^{\rm s}$ are acceptable for use in our EOS without
modification.  To determine the $\Theta_n^{\rm s}$ at any volume, we
first constructed a functional fit to the $\Theta_0^{\rm s}$ points, 
with the result
\begin{equation}
\Theta_0^{\rm s}(V) = 2852.69 + \frac{17\,319.9}{V} + 2.33667 \, V - 
                      633.858 \ln(V), 
\label{theta0}
\end{equation}
where $\Theta_0^{\rm s}$ is in K and $V$ is in $a_0^3$.  Then we noted
that according to the DFT results both $\Theta_1^{\rm s}$ and
$\Theta_2^{\rm s}$ approximately equal $e^{1/3}\,\Theta_0^{\rm s}$, so
we made the same approximation using Eq.\ (\ref{theta0}) for
$\Theta_0^{\rm s}$ to determine $\Theta_1^{\rm s}$ and $\Theta_2^{\rm
s}$ at any volume.  These functions are also shown in Figure
\ref{Thetas}.

\begin{figure}
\centering
\includegraphics[scale=0.75]{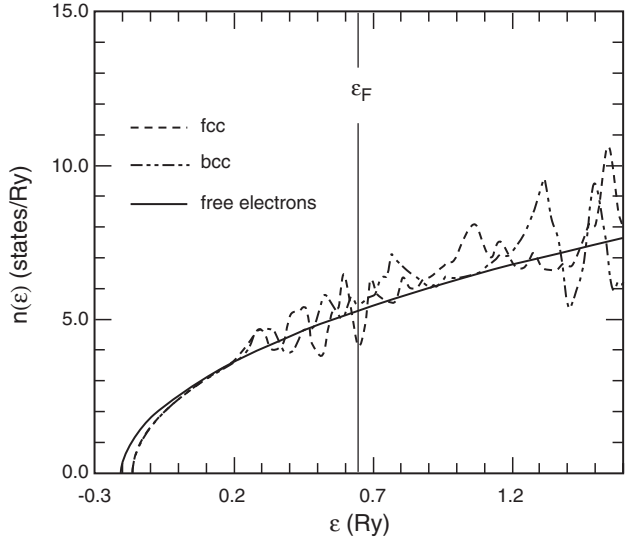}
\caption{$n^{\rm s}(\epsilon)$ for bcc and fcc Al at atomic volume
$112.0\ a_0^3$ from the calculations in \cite{straub}.  The free
electron $n^{\rm s}(\epsilon)$ at this volume and $Z=3$ from Eq.\
(\ref{freeel}) is shown for comparison. (From \cite{book}.)}
\label{dens}
\end{figure}

The DFT calculations also provided data on the electronic density of
states $n^{\rm s}(\epsilon)$.  Graphs of $n^{\rm s}(\epsilon)$ for fcc
and bcc Al at atomic volume $112.0\ a_0^3$ (corresponding to $P=0$ and
$T=295$ K) are shown in Figure \ref{dens}, along with the free
electron $n^{\rm s}(\epsilon)$, for which
\begin{equation} 
n^{\rm s}(\epsilon) = \sqrt{\frac{\epsilon}{\epsilon_F}}
                      \left(\frac{3Z}{2\epsilon_F}\right) 
\ \ \ \ \ {\rm and} \ \ \ \ \ 
\epsilon_F = \frac{\hbar^2}{2m_e} 
                 \left(\frac{3\pi^2Z}{V}\right)^{2/3},
\label{freeel}
\end{equation}
at $V = 112.0\ a_0^3$ and $Z=3$.
The Figure shows that the free electron model is a good approximation
for either crystal structure, for electronic excitations to around
$\frac{1}{2}$ Ry.  The same match, at a slightly poorer level of
approximation and for excitations to around $1$ Ry, is found at our
smallest atomic volume of $37\ a_0^3$.  For all volumes of our study
and temperatures up to $T_m$, the total electronic excitation
contribution to the energy, entropy, and pressure is at most $5\%$, so
the error introduced by using the free electron $n^{\rm s}(\epsilon)$
in our calculations is negligible.  Making this approximation, the
normalization condition from Eq.\ (\ref{findmu}) becomes
\begin{equation}
F_{1/2}(\beta\mu) = \frac{2}{3}(\beta\epsilon_F)^{3/2},
\end{equation}
which determines $\mu$ as a function of $V$ and $T$, and Eq.\
(\ref{Felgen}) for the free energy becomes
\begin{equation}
F_{\rm el}^{\rm s}(V,T) = Z\left( \mu - \frac{2}{3}\,kT\,\frac{F_{3/2}
(\beta\mu)}{F_{1/2}(\beta\mu)} - \frac{3}{5} \epsilon_F \right)
\label{Fel}
\end{equation}
where $Z=3$.  The $F_n(x)$ are the standard Fermi integrals; their
properties are discussed on pp.\ 332-334 of \cite{wilson} and their
values for various $n$ are tabulated in \cite{mcd,rhodes}.

\begin{figure}
\centering
\includegraphics[scale=0.75]{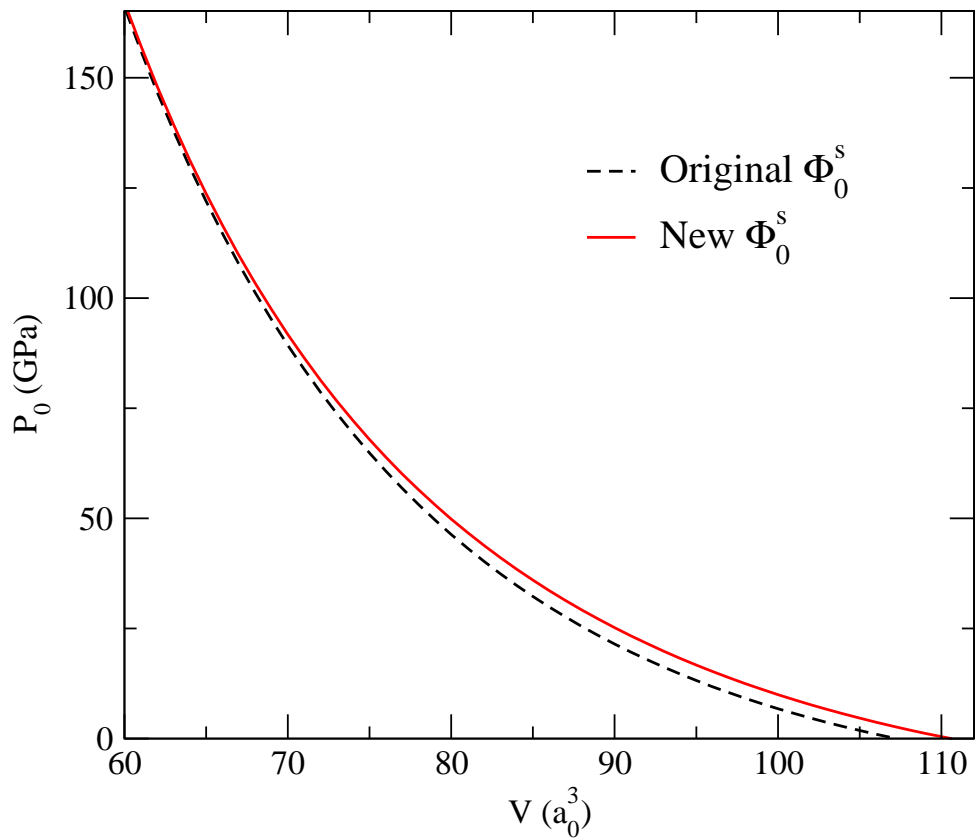}
\caption{The $T=0$ pressure-volume relations calculated using the
original $\Phi_0^{\rm s}$ and the new $\Phi_0^{\rm s}$ we constructed.
Notice how they differ in the vicinity of $V=110.6\ a_0^3$ but then
agree more closely at lower volumes.}
\label{phi0}
\end{figure}

The solid EOS that results from assembling all of these functions is
reliable over a large range of volumes and temperatures; however, it
is not in perfect agreement with the highly accurate experimental data
that are available at low pressures.  Specifically, experiments on Al
at $T=0$ and $P=0$ show that \cite{straub}
\begin{eqnarray}
& & V_0 = 110.6\ a_0^3, \ \ \ \ \ \ \ E_0 = -249\ {\rm mRy}, \nonumber \\
& & B_0 = 79.4 {\rm \ GPa}, \ \ \ \frac{dB_0}{dP} = 4.7,
\label{P0dat}
\end{eqnarray}
but the EOS yields $V_0 = 107.3\ a_0^3$, which is outside the
experimental error.  Therefore, we chose to make a small correction to
our purely theoretical free energy to agree with experiment.  These
quantities are obviously determined by the $T=0$ form of the free
energy, $F_0^{\rm s} = \Phi_0^{\rm s} + \frac{9}{8} k\Theta_1^{\rm s}$
(see Eq.\ (\ref{FlattlowT})); since $\Theta_1^{\rm s}$ is already in
excellent agreement with available experiment, we chose to modify only
$\Phi_0^{\rm s}$.  To proceed, we noted that the data determine $P_0$,
the $T=0$ pressure, in the vicinity of $V=V_0$ by the relation
\begin{eqnarray}
P_0(V) & = & P_0(V_0) + \left.\frac{dP_0}{dV}\right|_{V_0}(V-V_0) + 
        \frac{1}{2}\left.\frac{d^2P_0}{dV^2}\right|_{V_0}(V-V_0)^2 + 
        \cdots \nonumber \\
       & = & -\frac{B_0}{V_0}\,(V-V_0) + \frac{B_0}{2V_0^2}
              \left(1 + \frac{dB_0}{dP}\right)\,(V-V_0)^2 + \cdots 
\label{Pfromdat}
\end{eqnarray}
while at higher compressions we have no information to supplement the
electronic structure result; so we decided to construct a $\Phi_0^{\rm
s}$ that correctly reproduces Eq.\ (\ref{Pfromdat}) near $V_0$ but
smoothly interpolates to Eq.\ (\ref{firstphi}) at lower volumes.  To
do this, we computed $P_0$ at 10 volumes between $110\ a_0^3$ and
$111.25\ a_0^3$ using Eq.\ (\ref{Pfromdat}), and we also computed $P_0
= -\partial F_0^{\rm s} / \partial V$ using the above form for
$F_0^{\rm s}$, with $\Phi_0^{\rm s}$ from Eq.\ (\ref{firstphi}), at 23
volumes between $30\ a_0^3$ and $41\ a_0^3$.  We then performed a
least-squares fit to these points using an expression similar to the
Birch-Murnaghan form, but carried to a slightly higher order; after
integrating, adjusting the constant of integration to correctly match
$E_0$ from Eq.\ (\ref{P0dat}), and subtracting off $\frac{9}{8}k
\Theta_1^{\rm s}$, we had a new $\Phi_0^{\rm s}$ given by
\begin{eqnarray}
\Phi_0^{\rm s}(V) & = & -1.64615 \times 10^6 
                       + \frac{2.07608 \times 10^7}{V^{2/3}} 
                       - \frac{4.61515 \times 10^8}{V^{4/3}} \nonumber \\
                  &  & + \frac{5.71249 \times 10^9}{V^2}
                       - \frac{5.49998 \times 10^{10}}{V^{8/3}} 
                       + \frac{3.71978 \times 10^{11}}{V^{10/3}} \nonumber \\ 
                  &  & - \frac{1.66284 \times 10^{12}}{V^4}
                       + \frac{4.41118 \times 10^{12}}{V^{14/3}} 
                       - \frac{5.25064 \times 10^{12}}{V^{16/3}}  \nonumber \\
                  &  & + \frac{2.26789 \times 10^7}{V} 
                       - 220.716 \, V + 23\,6788 \ln(V)
\label{realphi}
\end{eqnarray}
where $\Phi_0^{\rm s}$ is in mRy and $V$ is in $a_0^3$.  This
$\Phi_0^{\rm s}$, which reproduces the data in Eq.\ (\ref{P0dat}) and
interpolates smoothly to the DFT curve at higher compressions, is what
we use in our EOS instead of Eq.\ (\ref{firstphi}).  The $T=0$
pressure-volume curves constructed using both the original and new
$\Phi_0^{\rm s}$ are shown in Figure \ref{phi0}.

Our choice of a Birch-Murnaghan-like form was dictated by the fact
that the Straub et al.\ result provides most of our information about
the shape of $\Phi_0^{\rm s}$; so our goal was to preserve that form
insofar as was possible, interpolating back to their result as quickly
as we could without introducing enough curvature to compromise
agreement with $dB_0/dP$.  This correction to $\Phi_0^{\rm s}$
constitutes a small change to the overall EOS; the effect of this
change on the Hugoniot will be considered in the next Section.  This
modification completes the full solid free energy, so we can now
consider the liquid.

For the liquid we need the same three functions that we needed for the
solid, and we must also consider the melting curve $T_m(P)$.  Having
chosen to use Eq.\ (\ref{FlatthighT}) for $F^{\rm s}_{\rm ph}$, we
certainly did the same for $F^l_{\rm ph}$, since the Hugoniot will
obviously enter the liquid region only at rather high temperatures;
thus we needed only $\Theta_0^l$ and $\Theta_2^l$.  From experiment we
know that Al is what is called in liquid dynamics a ``normal melting
element'' (the entropy of melting at constant density is approximately
$0.8\,k$ per atom), and we argue in \cite{liq} that $\Theta_0$ in
solid and liquid phases of such an element are approximately equal.
(Experimental and computational work supporting this conjecture are
also discussed in \cite{liq}.)  Thus we took the liquid to have the
same $\Theta_0$ as the solid.  It is also true that in the liquid, $T$
is typically much larger than $\Theta_2^{\rm s}$ (for example, in
liquid Al at normal density $T \geq 2\Theta_2^{\rm s}$), rendering the
first quantum correction to $F_{\rm ph}^l$ negligible (roughly $1\%$
at normal density), so even if $\Theta_2^l$ differs from
$\Theta_2^{\rm s}$ by $25\%$ or more, the impact on the phonon term
will be very small; therefore we also used the same $\Theta_2$ in the
liquid as in the solid.

Since the free electron model approximates the DFT result for $n^{\rm
s}(\epsilon)$ so well for both fcc and bcc Al (Figure \ref{dens}),
which correspond to two valleys in the many-body potential surface
with rather different structures, we also expect this model to be a
good approximation for $n^l(\epsilon)$, the density of states for the
structure characteristic of a liquid.  Since at all volumes and
temperatures up to $5\,T_m$ (the relevance of this number will appear
below), the electronic contribution to the thermodynamic functions
does not exceed $25\%$, the error introduced by the free electron
model is still acceptable.

\begin{figure}
\centering
\includegraphics[scale=0.75]{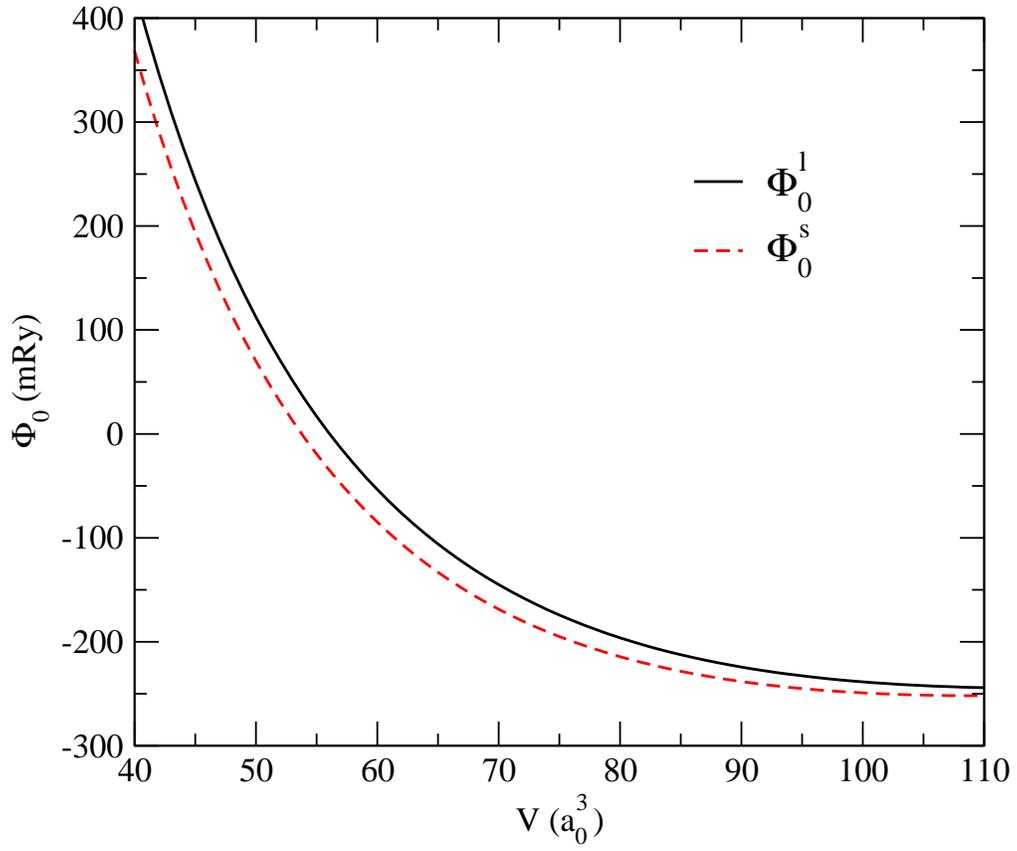}
\caption{$\Phi_0^l$ determined by matching the liquid and solid Gibbs
free energies along the melt curve. $\Phi_0^{\rm s}$ is also shown for
comparison.}
\label{phi0l}
\end{figure}

We fixed the only remaining term in Eq.\ (\ref{realFl}), $\Phi_0^l$, by
the requirement that the Gibbs free energies of the solid and liquid
match along the Al melting curve.  Boehler and Ross \cite{boe}
suggested that
\begin{equation}
T_m(P) = 933.45\ {\rm K}\left(\frac{P}{6.049\ {\rm GPa}} + 1\right)^{0.531}
\label{TP}
\end{equation}
on the basis of their experimental work up to 80 GPa (0.8 Mbar), and
experiments by McQueen et al.\ \cite{mcq} and H\"{a}nstr\"{o}m and
Lazor \cite{hans} and theoretical work by P\'{e}lissier \cite{peli}
suggest that this curve continues to be accurate up to 200 GPa.  In
the absence of evidence to the contrary, we took Eq.\ (\ref{TP}) to be
valid to higher pressures as needed.  (As we will see later, our EOS
will assume Eq.\ (\ref{TP}) no higher than 400 GPa.)  We computed
$\Phi_0^l$ as follows: We made a guess for $\Phi_0^l$ not very
different from $\Phi_0^{\rm s}$, and then we used it and Eq.\
(\ref{TP}) to calculate the difference between the two Gibbs free
energies,
\begin{equation}
\Delta G(P) = G^{\rm s}(P, T_m(P)) - G^l(P, T_m(P)),
\label{DeltaG}
\end{equation}
at several hundred values of $P$ over the entire pressure range
considered in this study.  We also calculated the liquid melt volume
$V_m^l(P)$ at each $P$.  If the rms average of Eq.\ (\ref{DeltaG})
over all calculated points was not sufficiently small, we used the
following easily verified fact: To first order, a small change $\delta
\Phi_0^l$ produces a small change $\delta G^l(P, T_m(P))$ given by
$\delta G^l(P, T_m(P)) = \delta \Phi_0(V_m^l(P))$.  Thus we performed
the substitution
\begin{equation}
\Phi_0^l(V) \rightarrow \Phi_0^l(V) + \Delta G(P_m^l(V)),
\end{equation}
where $\Delta G$ was computed by Eq.\ (\ref{DeltaG}) and $P_m^l(V)$ is
the inverse of $V_m^l(P)$, and calculated Eq.\ (\ref{DeltaG}) again.
We iterated until the rms average was sufficiently small (less than
$0.001$ mRy in our case), giving us the needed $\Phi_0^l$, which is
shown in Figure \ref{phi0l} along with $\Phi_0^{\rm s}$.  We recorded
$\Phi_0^l$ as a list of points, and we did not create a functional fit
for it; instead we used an interpolator to calculate it and its
derivative when needed.

It is at this point that the existence of other solid phases in Al,
discussed earlier, affects the EOS of the liquid.  It is likely that
the liquid region borders the fcc crystal only over part of its
boundary, beyond which the liquid borders the bcc region or other
solid phases.  In such a case, at sufficiently high pressures we
should use the free energy appropriate for that solid phase, not the
fcc free energy, in Eq.\ (\ref{DeltaG}).  This suggests that
$\Phi_0^l$ may become inaccurate beyond densities in the neighborhood
of 6 g/cm$^3$, where the $T=0$ fcc-bcc phase transition occurs.  We
will take this fact into consideration when we discuss the limits of
applicability of the EOS below.

\begin{figure}
\centering
\includegraphics[scale=0.75]{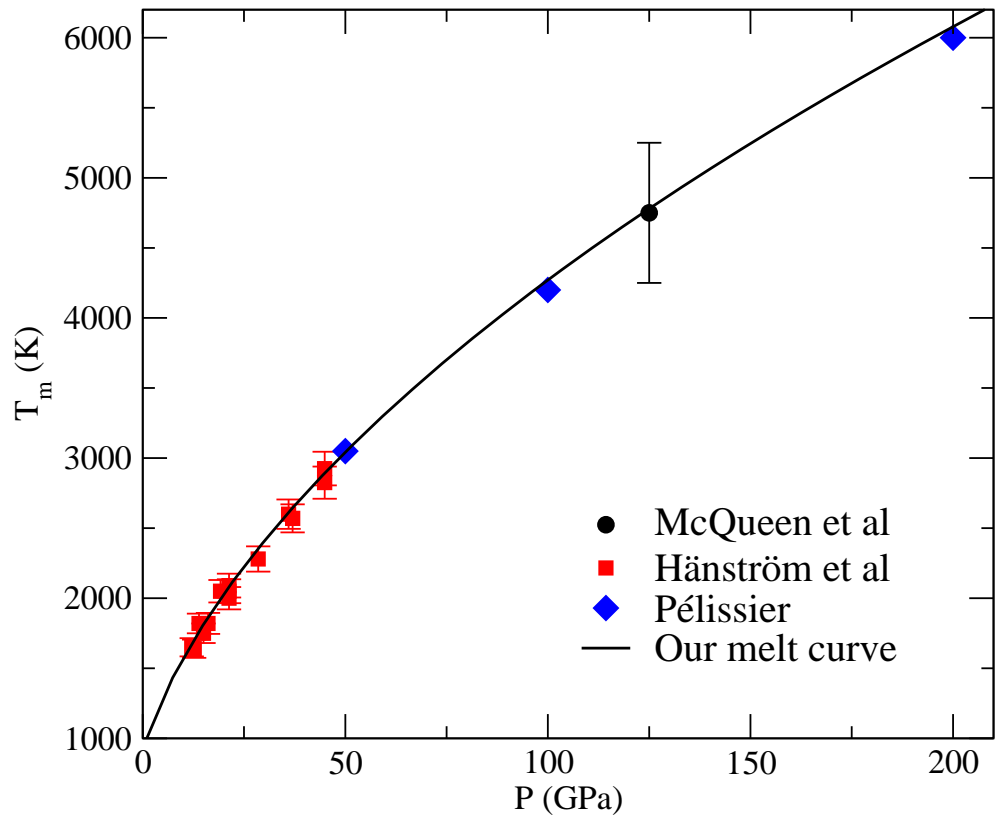}
\caption{The melt curve $T_m(P)$ computed from our full solid and
liquid EOS (which reproduces the Boehler-Ross curve, Eq.\ (\ref{TP})),
the experimental data from \cite{mcq,hans}, and points from the
theoretical curve in \cite{peli}.}
\label{Tm}
\end{figure}

Once we had the full solid and liquid EOS, we then solved Eq.\
(\ref{Gibbsmatch}) directly to compute $T_m(P)$, verifying that we had
reproduced the Boehler-Ross curve; our result is shown in Figure
\ref{Tm}, together with the data from \cite{mcq,hans} and some
points from P\'{e}lissier's theoretical curve.  (According to
\cite{mcq}, their data point at 125 GPa marks the onset of melting
along the Hugoniot.  We will comment on this in the next Subsection.)

Now that we have the full two-phase EOS, it is profitable to compare
our work with another EOS for Al, due to Moriarty et al.\ \cite{mor}.
These authors also use a full lattice dynamics treatment of the
crystal phonons, although they calculate their $g^{\rm s}(\omega)$ two
separate ways, using both Moriarty's Generalized Pseudopotential
Theory (GPT) and a local pseudopotential model with parameters chosen
to match solid-phase EOS data.  We strongly prefer to rely on DFT
results, as we believe DFT has reached such a stage of maturity that
it more accurately captures the physics contained in the real
Hamiltonian of the system, which as we have emphasized we believe to
be understood in sufficient depth that it should underlie all of our
work.  Second, in their treatment of the liquid phase Moriarty et al.\
rely on fluid variational theory, described in detail in \cite{ash},
to compute the least upper bound to the ``real'' liquid free energy
(from a liquid Hamiltonian based on pseudopotentials) that can be
obtained from the free energy of a reference system; Moriarty et al.\
investigate hard-sphere, soft-sphere, and one-component plasma
reference systems before settling on the soft-sphere system as
providing the best bound.  Based on the investigations summarized in
\cite{liq}, we claim that we have the actual Hamiltonian of the liquid
itself, not a Hamiltonian based on pseudopotential theory;
furthermore, this Hamiltonian decomposes naturally into a dominant
term that produces a free energy that can be used directly (instead of
requiring approximation by the free energy of a reference system) and
remaining terms whose contributions to the free energy are known to be
small (see the Introduction).  The same point we made above for the
solid phase applies; we argue that it is a better strategy in
developing EOS to try to understand the true Hamiltonian of the
system, and then to use it when doing statistical mechanics.  Almost
inevitably, one must make approximations (which we certainly have done
here), but we believe we are in a better position to understand and
improve upon them if the physical foundation of the EOS is as solid as
we can make it.

Finally, let us make some conservative estimates of the range of
applicability of this EOS.  Any limits will arise from two sources:
the validity of the approximation that $F^l_{\rm ab}$ is negligible in
the liquid (see the Introduction), and the limited ranges over which
the functions $\Phi_0(V)$, $g(\omega)$, $n(\epsilon)$, and $T_m(P)$ are
known.  Let's consider each in turn.

(1) We know from experiment that $F^l_{\rm ab}$ is negligible when $T
\leq 3\,T_m$ (again, see the Introduction), and judging from trends in
the data we suspect $F^l_{\rm ab}$ will still be small up to $T
\approx 5\,T_m$, but clearly this term must become relevant as the
nuclear motion becomes more gaslike.  Thus we shall take care with any
data at $\rho$ and $T$ such that $T$ approaches or exceeds
$5\,T_m(\rho)$.

(2) At densities below approximately 6 g/cm$^3$, we are confident that
the solid is in the fcc phase, and the liquid free energy is based on
this phase, so we trust the full EOS here.  At higher densities, we
must be more circumspect; the solid may have undergone a phase
transition to bcc, and the liquid EOS at this density may be based on
the wrong solid free energy.  Further, as we have indicated earlier,
Eq.\ (\ref{TP}) for the melt curve has received independent support
only up to 200 GPa, so we must be cautious with the liquid EOS in
regions beyond this point.  We decided to be brave and accept the melt
curve as valid up to 400 GPa; this corresponds to a liquid density of
$6.15$ g/cm$^3$, and since this is not far from the probable location
of the solid fcc-bcc transition, we take it as the density limit of
our EOS.  (Even if we did not have this concern, we would be
restricted to densities below $8.17$ g/cm$^3$, where electronic
structure results are available.)  Also, the free electron
approximation to $n^{\rm s}(\epsilon)$ has been validated only for
$\epsilon - \epsilon_F$ up to 1/2 Ry, or 6.8 eV, at low compression
and 1 Ry, or 13.6 eV, at high compression, but at higher temperatures
the electronic energy and entropy are sensitive to the details of
$n^{\rm s}(\epsilon)$ to energies above these limits.  We estimated
the values of $T$ that begin to probe the unvalidated region of
$n^{\rm s}(\epsilon)$ (roughly $3kT = \epsilon - \epsilon_F$), and we
found that over our valid density range the $T = 5\,T_m$ limit always
took precedence.  Hence this limit is not relevant for us, but we
mention it for completeness, as it may become a concern if the EOS is
extended to higher densities.

\begin{figure}
\centering
\includegraphics[scale=0.75]{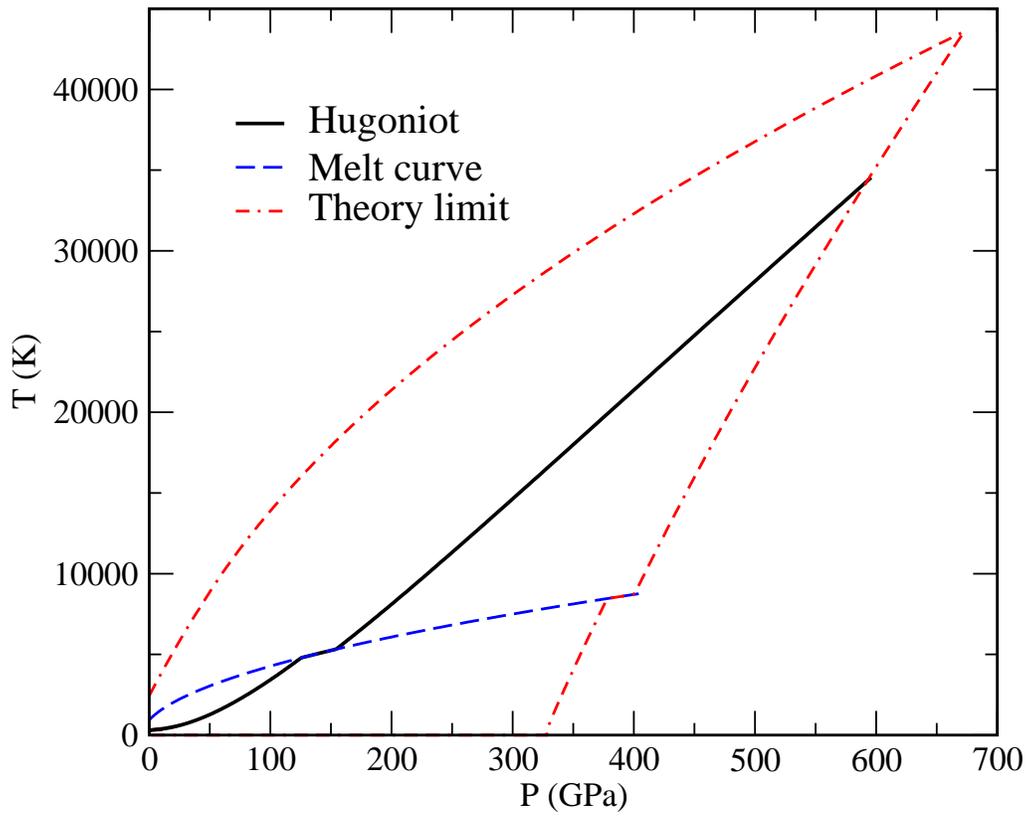}
\caption{The limits of our EOS, the melt curve, and the Hugoniot.}
\label{TPlim}
\end{figure}

\begin{figure}
\centering
\includegraphics[scale=0.75]{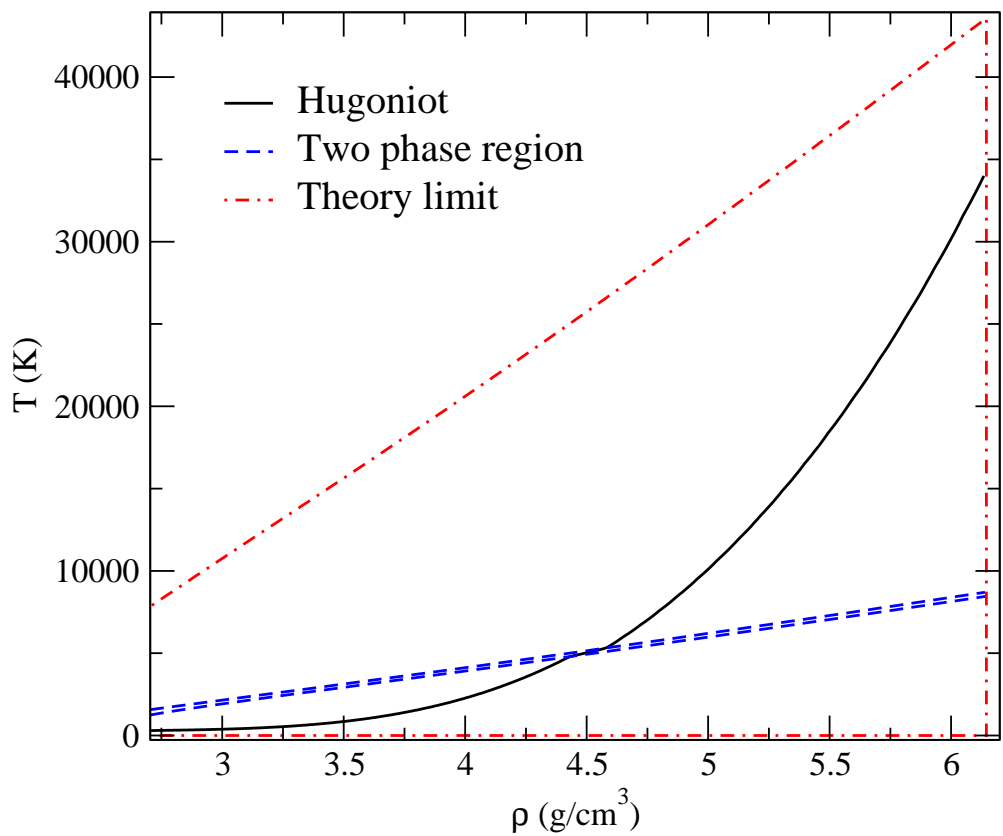}
\caption{The limits of our EOS, the two-phase region (solid below the
region, liquid above), and the Hugoniot.}
\label{Trholim}
\end{figure}

Figure \ref{TPlim} shows the limits $\rho \leq 6.15$ g/cm$^3$ and $T
\leq 5\,T_m$ of the EOS in $T-P$ space, together with the melt curve
and the Hugoniot (see the next Subsection), while Figure \ref{Trholim}
shows the same three features in $T-\rho$ space.
In this Figure, the melt curve becomes a two-phase region, which
we will consider in more detail in the next Subsection.
We are confident that this EOS is valid within these limits, but we
don't know how far beyond them the inaccuracies begin to appear; thus
we will not be shy about considering data not too far outside this
range.

\subsection{Comparison with Hugoniot data}
\label{hug}

If a shock wave travels at speed $u_s$ through a sample of material,
accelerating its particles from rest to speed $u_p$ and changing its
density, atomic volume, pressure, and internal energy per atom from
$\rho_0$, $V_0$, $P_0$, and $E_0$ to $\rho$, $V$, $P$, and $E$, then
(assuming thermal equilibrium before and after the shock) these
quantities must satisfy the Rankine-Hugoniot relations,
\begin{eqnarray}
\rho (u_s - u_p) & = & \rho_0 u_s \nonumber \\
P - P_0 & = & \rho_0 u_s u_p \nonumber \\
E - E_0 & = & \frac{1}{2} (P_0 + P)(V_0 - V),  
\label{ranhug}
\end{eqnarray}
derived from considerations of mass, momentum, and energy
conservation.  (It is assumed that the wave is steady and strength
effects are negligible.)  

\begin{figure}
\centering
\includegraphics[scale=0.75]{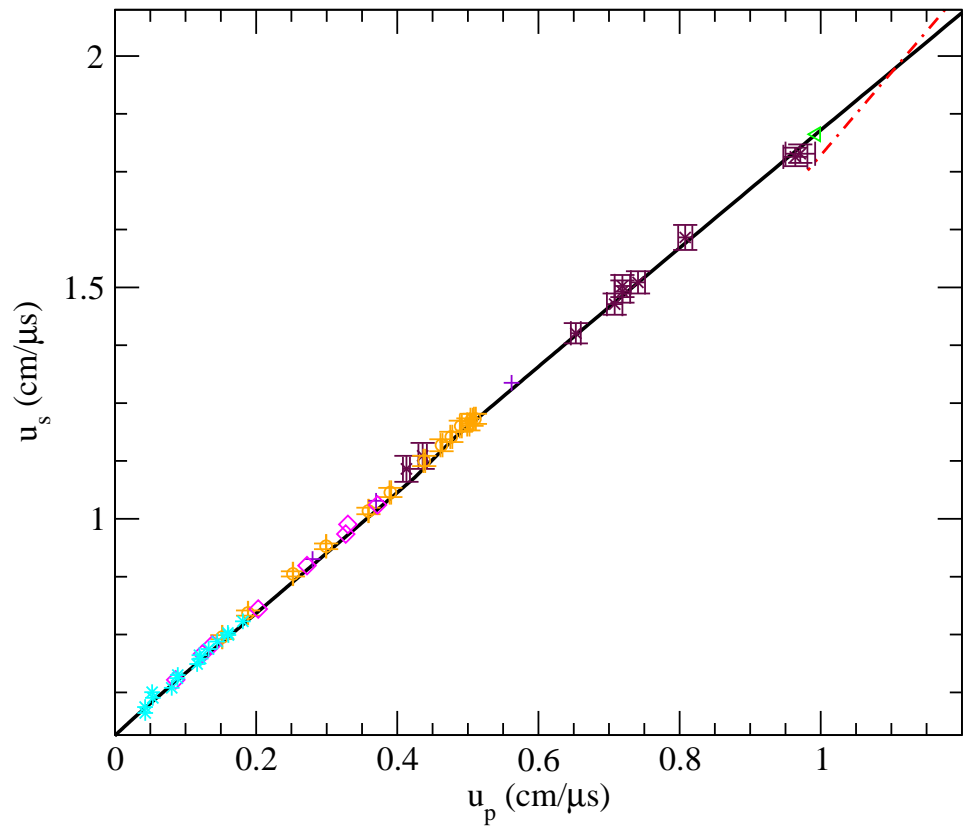}
\caption{The $u_s$-$u_p$ Hugoniot for Al predicted by our EOS together
with experimental data from \cite{alt,kor,LASL,alt2,mitnell,knud}.
The intersection of the Hugoniot with the limit of validity of the EOS
(dot-dash line) is also indicated.}
\label{usup}
\end{figure}
\begin{figure}
\centering
\includegraphics[scale=0.75]{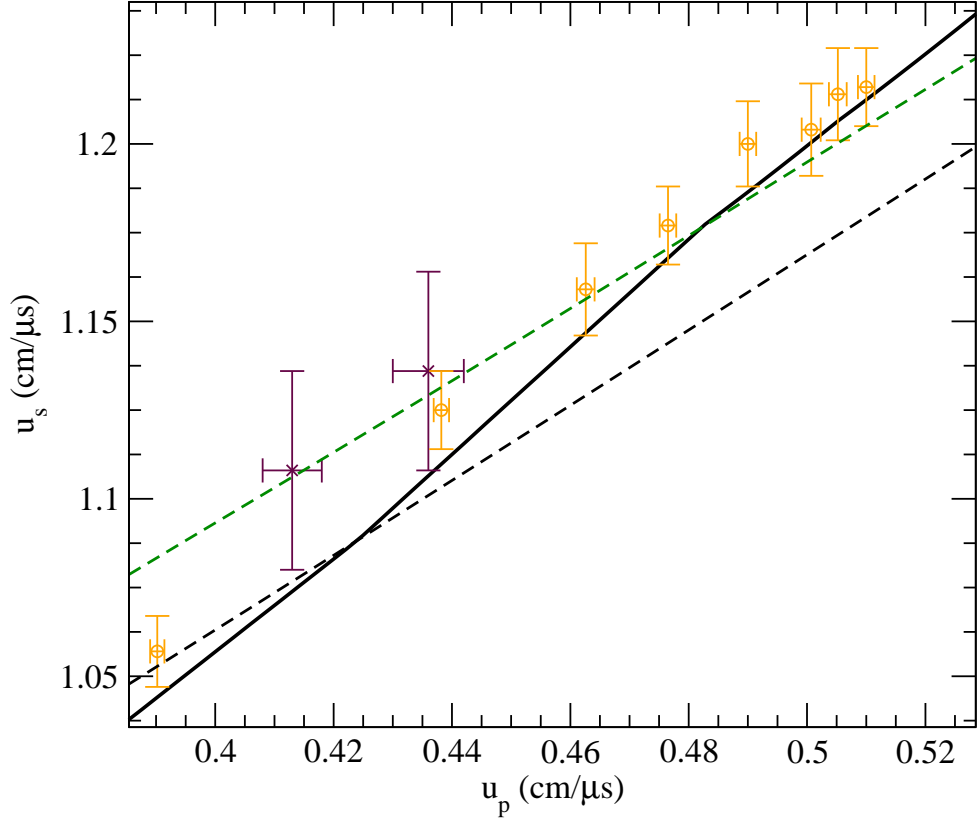}
\caption{The $u_s$-$u_p$ Hugoniot in the intermediate-$P$ region,
including intersections with the phase boundaries, with data from
\cite{mitnell,knud}.}
\label{usupzoomtwoph}
\end{figure}
\begin{figure}
\centering
\includegraphics[scale=0.75]{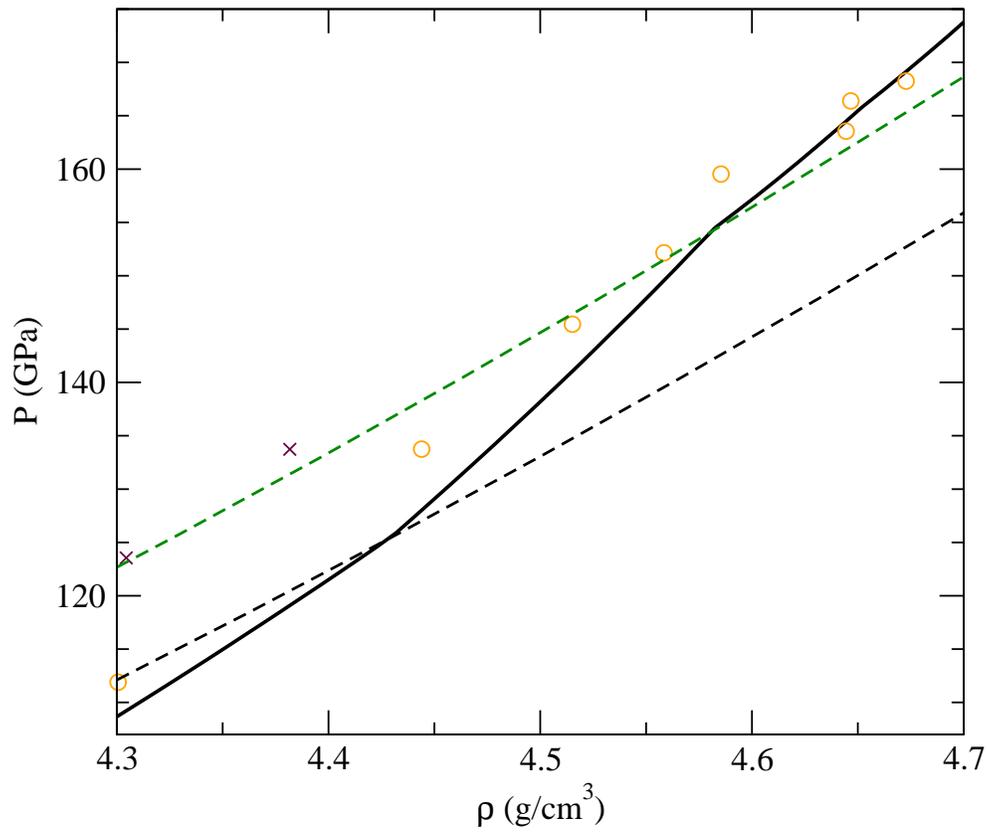}
\caption{The $P$-$\rho$ Hugoniot in the intermediate-$P$ region,
including intersections with the phase boundaries, with data from
\cite{mitnell,knud}.}
\label{Prhozoomtwoph}
\end{figure}

By solving these equations together with the EOS, which relates $P$,
$V$, and $E$, we can determine the Hugoniot, the curve of all possible
end states of the shocked material. 
\begin{figure}
\centering
\includegraphics[scale=0.75]{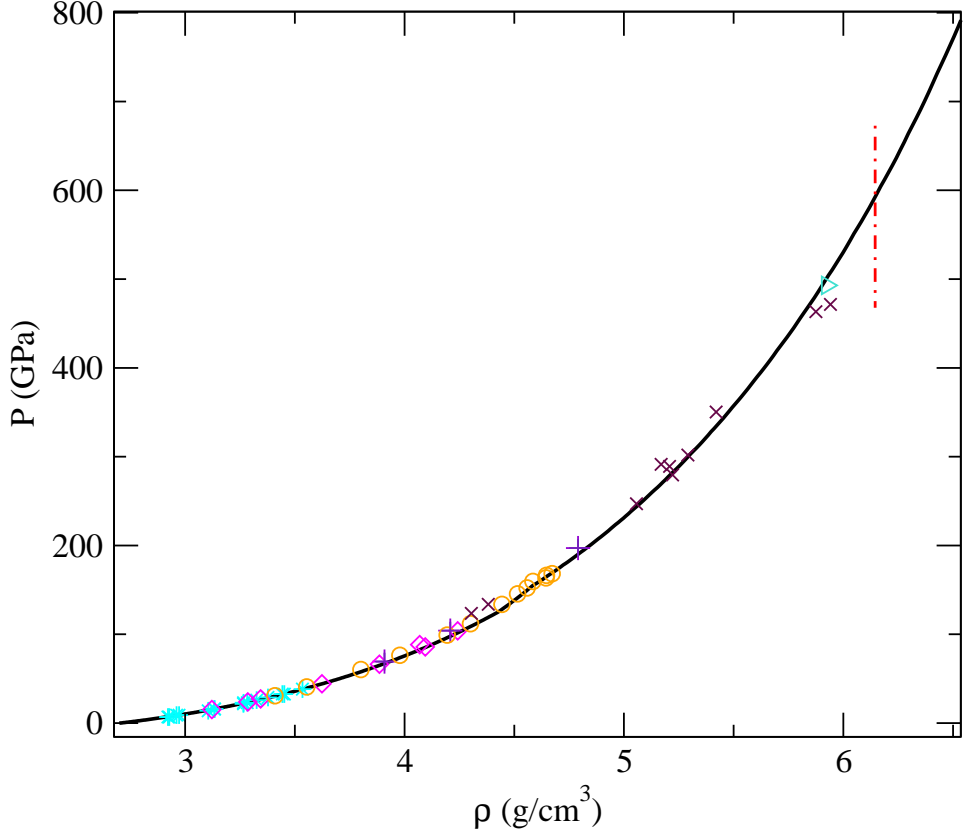}
\caption{The $P$-$\rho$ Hugoniot for Al predicted by our EOS together
with experimental data from \cite{alt,kor,LASL,alt2,mitnell,knud}.
The intersection of the Hugoniot with the limit of validity of the EOS
(dot-dash line) is also indicated.}
\label{Prho}
\end{figure}
We used our EOS and Eqs.\ (\ref{ranhug}) to compute $u_s$ as a
function of $u_p$ and $P$ as a function of $\rho$ along the Al
Hugoniot; the results are shown in Figures \ref{usup} and \ref{Prho}
along with the intersection of the Hugoniot with the limit of validity
of the EOS.
\begin{figure}
\centering
\includegraphics[scale=0.75]{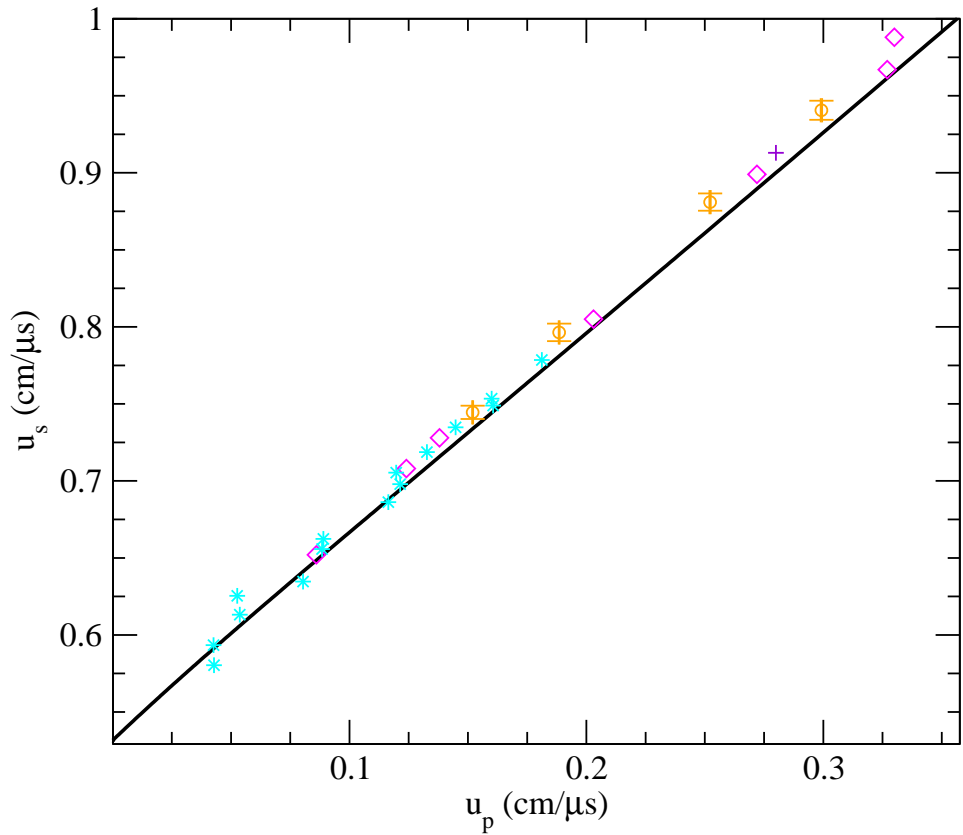}
\caption{The $u_s$-$u_p$ Hugoniot in the low-$P$ region, with data
from \cite{alt,LASL,alt2,mitnell}. The $u_p$ error bars on the circles
\cite{mitnell} appear as slightly broadened vertical lines.}
\label{usupzoomlowP}
\end{figure}
Hugoniot data from several sources
\cite{alt,kor,LASL,alt2,mitnell,knud} are also included.
\begin{figure}
\centering
\includegraphics[scale=0.75]{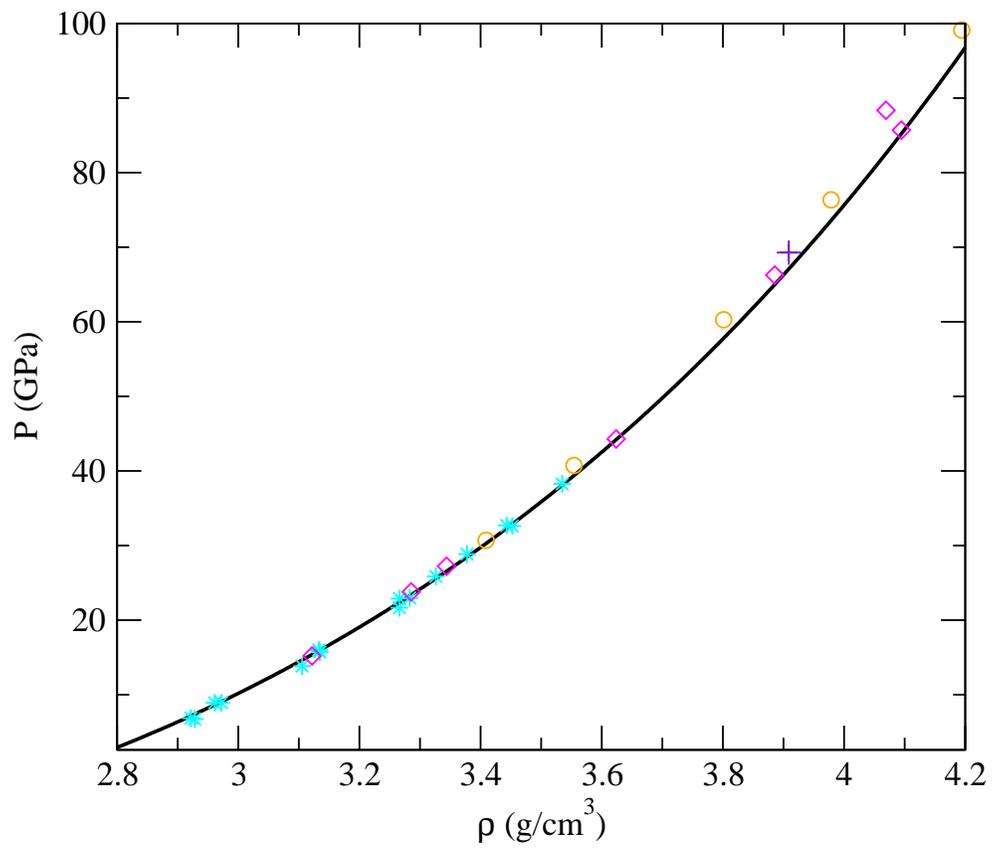}
\caption{The $P$-$\rho$ Hugoniot in the low-$P$ region, with data from
\cite{alt,LASL,alt2,mitnell}.}
\label{PrhozoomlowP}
\end{figure}
The low-pressure region of the Hugoniot is highlighted in Figures
\ref{usupzoomlowP} and \ref{PrhozoomlowP}, and the
intermediate-pressure region, including the intersections with the
phase boundaries, is shown in Figures \ref{usupzoomtwoph} and 
\ref{Prhozoomtwoph}.

Three important considerations in selecting which data to include are
(1) the initial densities of the samples, (2) the quality of the
experimental technique, and (3) whether the measurements were absolute
or relative.  
All of the available data were taken using Al alloys
with densities that differ from the known pure metal value of $2.70$
g/cm$^3$ (predicted correctly by our EOS); some alloys are as close as
$2.71$ g/cm$^3$ while others differ much more.  
Since Hugoniots in general are quite sensitive to the initial density,
we chose to compare only with the data for which $\rho_0$ clustered
around $2.71$ g/cm$^3$.  (Thus we used only one data point from
\cite{kor}, which mainly concerns porous materials.  All of the data
from the other references were used.)  We also avoided sources which
gathered data using unusual shock wave geometries (such as
\cite{skid}), and we also chose not to use the results of indirect or
relative measurements, such as \cite{alt3,trun,glus}, preferring to
rely on the absolute measurements that are available.  Finally, we did
not use the few data points available (primarily nuclear-driven) that
lie very far beyond the limits of applicability of our EOS (but see
below).

The theoretical Hugoniot compares well with both the $u_s$-$u_p$ and the
$P$-$\rho$ data all the way up to the predicted limit of its validity,
at approximately $500$ GPa ($5$ Mbar).  More specifically, theory
agrees with experiment at $P \stackrel{<}{\sim} 40$ GPa (Figures
\ref{usupzoomlowP} and \ref{PrhozoomlowP}); at 40-125 GPa, theory
falls below the experimental error bars by around 1\% at most (Figures
\ref{usupzoomlowP} through \ref{Prhozoomtwoph}); and theory again lies
within the experimental error bars through the liquid phase (Figures
\ref{usup}, \ref{Prho}, \ref{usupzoomtwoph}, and \ref{Prhozoomtwoph}).
(We recall that given percentage errors in $u_s$ and $u_p$ correspond to
roughly the same percentage errors in the $P$-$\rho$ plane.)  The
presence of theoretical error only in the solid phase is likely due to
strength effects, which are present in the solid but not in the
liquid, and which are neglected in our Hugoniot calculations.
Furthermore, as Figure \ref{Prhozoomtwoph} shows, we predict that the
Hugoniot crosses the two-phase region between $\rho = 4.43$ g/cm$^3$
and $\rho = 4.58$ g/cm$^3$, corresponding to a range in $P$ from 126
to 156 GPa; this agrees very well with \cite{mcq}, in which melting
was found to occur between 125 and 150 GPa.  (We note that \cite{mcq}
used Al 2024, an alloy whose density is sufficiently different from
pure Al that we did not use Hugoniot data taken with that alloy in our
Figures.  We consider their melting results because, as we saw in the
previous Subsection, their data are consistent with other experiments
that did use pure Al.)  The correction to $\Phi_0^{\rm s}$ from the
previous Subsection, shown in Figure \ref{phi0}, shifts the Hugoniot
at pressures below 30 GPa, bringing it into excellent agreement with
experiment, while at pressures above 60 GPa or so, the effect on the
Hugoniot is insignificant.

We have also compared our results with data just beyond the EOS limits
of validity; the points in \cite{sim} that match our initial density
(one of which is a reanalysis of the single point in \cite{vol}),
lying at about 10 Mbar, fall noticeably below our Hugoniot, and their
consistency with the very-high-pressure points of Ragan
\cite{rag1,rag2} strongly suggests that they represent the true
Hugoniot, which thus falls beneath our prediction at higher pressures.
Possible errors in our EOS at such densities include, in what we
estimate to be decreasing order of importance, (1) the shift from the
fcc to the bcc crystal, with a corresponding change in $\Phi_0^l$ as
discussed in the last Subsection, (2) deviations in the melt curve
$T_m(P)$ from the Boehler-Ross form at higher pressures (the densities
of the points in \cite{sim} correspond to melt pressures around 620
GPa according to our EOS), (3) the fact that at such high $T$ the EOS
is probing the high-energy region of $n^{\rm s}(\epsilon)$, and (4)
the neglect of the anharmonic and boundary contributions to the liquid
EOS ($T$ is only slightly below $5\,T_m$ at these points according to
our EOS).

\section{Conclusions}
\label{concl}

Drawing upon theory developed in \cite{book}, we have described a
framework for constructing EOS for elemental solids and liquids, and
we have discussed experimental and theoretical results indicating that
the framework remains highly accurate at low pressures when certain
small effects (anharmonicity, boundaries, electron-phonon
interactions) are neglected.  After displaying the resulting formulas
for the Helmholtz free energy, we considered the information one needs
to evaluate them, and we discussed the combination of experiment and
theoretical work that could be used to get this information.  Finally,
as an illustration we constructed an EOS for Al, established its range
of validity based on the inputs to the EOS, and compared it with
Hugoniot data to 5 Mbar; our EOS matched the data to the accuracy we
expected based on the low-pressure results.

We consider the primary advantage of this method to lie in the fact
that it incorporates into the decomposition of the Hamiltonian a great
deal of accumulated knowledge of condensed matter physics both for the
solid and liquid phases (for example, the fact that the electronic
ground state energy is the most appropriate potential for the nuclear
motion).  If we have indeed captured the correct physics (and we
expect no new physics to enter until the relativistic domain), then
the EOS should have the right functional form, which means that if it
is shown to agree with available data, then we have reason to believe
that it will be equally accurate in regions where no data are
available; and making predictions where we have no data is the point
of having an EOS to begin with.  Furthermore, the better the
foundation we can build, the better our position for intelligently
investigating and controlling our approximations.

This discussion bears on the second goal of this paper, which was to
learn whether the approximation of neglecting anharmonic, boundary,
and electron-phonon effects remains useful at higher pressures.  We
already knew, as discussed in the Introduction, that at low $P$ the
anharmonic and electron-phonon terms are small, and we found this from
direct calculations; we also knew that for several elements, over a
range of $T$, at low $P$ the approximations in question yielded
thermal energies and entropies that disagreed with experiments by
$5\%$ at most.  Our work here has shown that for one material at much
higher $P$ the approximations yield results that match data along a
single curve, the Hugoniot, to comparable accuracy.  Based on our
arguments above, that the EOS incorporates the correct physics and is
thus of the correct functional form, we claim to have shown that this
Al EOS is trustworthy throughout its range of validity, for all $T$
and $P$.

The main disadvantage of this method is that it relies on many inputs
($\Phi_0$, $g(\omega)$, and $n(\epsilon)$ for each phase) which may be
available only over limited ranges, and each of these limits also
restricts the range of validity of the EOS.  Our Al example amply
illustrates this problem; with a compression range from a little under
one to just over two, and a temperature range that reaches only to
slightly under 4 eV, this EOS is inadequate for many applications at
the national laboratories.  We argue, however, that this problem does
not indicate a deficiency in the approach; it only underscores the
need for many more DFT calculations of these quantities for more
materials with ever greater accuracy over ever larger ranges.

In the meantime, though, we would like to be able to say something
about elemental solid and liquid EOS at higher compressions.  We do
know that as density increases, the electrons come to dominate the
free energy, and it is also known that TFD correctly describes the
electrons in the limit of infinite density.  This suggests the
following possibility: Construct an EOS using the present techniques
to compressions as high as the available experimental or DFT results
allow, and then interpolate between these results and the predictions
of TFD for higher compressions.  This raises an important question: At
what pressures does TFD begin to become accurate?  Conventional
wisdom, usually traced back to Feynman et al.\ \cite{feyn}, has held
that TFD becomes reliable starting at $P \approx 10$ Mbar, but other
work \cite{boett} suggests that TFD (or TF in their case, but TF and
TFD converge at high pressures) deviates noticeably from electronic
structure results until 100 Mbar at least.  This suggests that the
pressure threshold at which TFD is trustworthy has not yet been
adequately established; it would be of great interest to settle this
question more definitively.

\begin{center} \bf \Large Acknowledgment \end{center}

This work was supported by the U.\ S.\ DOE through contract W-7405-ENG-36.

\end{document}